\documentclass[aps,prl,twocolumn,amsmath,amssymb,showpacs,superscriptaddress,notitlepage,longbibliography]{revtex4-1}
\pdfoutput=1
\usepackage[colorlinks=true,linkcolor=blue,anchorcolor=red,citecolor=blue, urlcolor=blue]{hyperref}
\usepackage{bm}
\usepackage{graphicx}
\usepackage{color}

\UseRawInputEncoding
\begin{document}

\title{Koopmans' theorem as the mechanism of nearly gapless surface states in self-doped magnetic topological insulators}

\author{Weizhao Chen}
\affiliation{Shenzhen Institute for Quantum Science and Engineering and Department of Physics, Southern University of Science and Technology, Shenzhen 518055, China}
\author{Yufei Zhao}
\affiliation{Shenzhen Institute for Quantum Science and Engineering and Department of Physics, Southern University of Science and Technology, Shenzhen 518055, China}
\author{Qiushi Yao}
\affiliation{Shenzhen Institute for Quantum Science and Engineering and Department of Physics, Southern University of Science and Technology, Shenzhen 518055, China}
\author{Jing Zhang}
\affiliation{Shenzhen Institute for Quantum Science and Engineering and Department of Physics, Southern University of Science and Technology, Shenzhen 518055, China}
\author{Qihang Liu}
\email{Corresponding Author: liuqh@sustech.edu.cn}
\affiliation{Shenzhen Institute for Quantum Science and Engineering and Department of Physics, Southern University of Science and Technology, Shenzhen 518055, China}
\affiliation{Guangdong Provincial Key Laboratory of Computational Science and Material Design, Southern University of Science and Technology, Shenzhen 518055, China}
\affiliation{Shenzhen Key Laboratory of for Advanced Quantum Functional Materials and Devices, Southern University of Science and Technology, Shenzhen 518055, China}

\date{\today}

\begin{abstract}
The magnetization-induced gap at the surface state is widely believed as the kernel of magnetic topological insulators (MTIs) because of its relevance to various topological phenomena, such as the quantum anomalous Hall effect and the axion insulator phase. However, if the magnetic gap exists in an intrinsic MTI, such as MnBi$_{2}$Te$_{4}$, still remains elusive, with significant discrepancies between theoretical predictions and various experimental observations. Here, including the previously overlooked self-doping in real MTIs, we find that in general a doped MTI prefers a ground state with a gapless surface state. We use a simple model based on Koopmans' theorem to elucidate the mechanism and further demonstrated it in self-doped MnBi$_{2}$Te$_{4}$/(Bi$_{2}$Te$_{3}$)$_{n}$ family through first-principles calculations. Our work shed light on the design principles of MTIs with magnetic gaps by revealing the critical role of doping effects in understanding the delicate interplay between magnetism and topology.

\end{abstract}

\maketitle

\emph{Introduction}.\rule[3pt]{0.4cm}{0.02em}Magnetic topological insulator (MTI) enables the interplay between magnetism and topological electronic structure, and thus is an ideal platform to realize exotic topological quantum phenomena \cite{Ref.1,Ref.2}. Typically, the magnetic moments open a band gap of 2$|\emph{M}|$ in an otherwise gapless topological surface state (TSS), described by the surface Hamiltonian $ \emph{v}_{_{F}}(\emph{k}_{x}\sigma_{y}-\emph{k}_{y}\sigma_{x})+\emph{M}\sigma_{z}$, where the first two terms present a linear Dirac cone and $\emph{M}$ the effective Zeeman field along the \emph{z} direction. Such a gap is essential for a MTI because it carries 1/2 topological charge, which is the manifestation of the quantized bulk topological magnetoelectric coupling. In addition, as the magnetization gap overwhelms the hybridization gap in a two-dimensional (2D) MTI slab, the quantum anomalous Hall effect (QAHE) can be observed \cite{Ref.3,Ref.4,Ref.5}.

In recent years, the MnBi$_{2}$Te$_{4}$-family compounds were found as ideal intrinsic MTI materials with theoretically-predicted sizable TSS gap of 80$\sim$100 meV \cite{Ref.6,Ref.7,Ref.8}. Surprisingly, while the surface gap of MnBi$_{2}$Te$_{4}$ was reported in some early reports \cite{Ref.6,Ref.9}, subsequent angle-resolved photoemission spectroscopy (ARPES) measurements \cite{Ref.10,Ref.11,Ref.12,Ref.13,Ref.14,Ref.15,Ref.16,Ref.17} had observed nearly perfect Dirac cone at its (0001) surface, robust across the critical temperature. Several hypotheses had been exposed for this discrepancy, focusing on the reconstruction of magnetic or geometric configuration at the surface, such as the in-plane or paramagnetic spin reorientation, surface relaxation of the top van der Waals (vdW) layer, etc \cite{Ref.10,Ref.12,Ref.18,Ref.19,Ref.20,Ref.21}. However, all these scenarios are phenomenological and specific to a gapless reconfiguration, lacking the evidence why such reconfiguration would occur. Moreover, while the nearly gapless TSS was also observed at the MnBi$_{2}$Te$_{4}$-termination of MnBi$_{4}$Te$_{7}$ \cite{Ref.10,Ref.15,Ref.22,Ref.23,Ref.24,Ref.25}, a magnetic gap about 28 meV was verified at the MnBi$_{2}$Te$_{4}$-termination of MnBi$_{8}$Te$_{13}$ \cite{Ref.25}. Therefore, a systematic theory that reconciles various experimental observations is desirable for understanding the nature of the magnetic gap in real MTI samples.

Here, skipping specific speculations of reconfiguration (magnetic or geometric) that would lead to gapless TSS, we consider a top-down question, i.e., if the tendency of the gapless TSS is representative for a group of MTIs with certain features, or material dependent to MnBi$_{2}$Te$_{4}$. We take into account an important experimental fact, yet overlooked by the previous scenarios, that the as-grown MTI samples are usually self-doped, such as the \emph{n}-type doped MnBi$_{2}$Te$_{4}$ \cite{Ref.26,Ref.27,Ref.28}. A simple model derived from the generalized Koopmans' theorem reveals that, regardless of \emph{n}-type or \emph{p}-type, the self-doped MTI with enough doping concentration prefers the gapless TSS. If the gain of the single-particle energy eigenvalues of the doped electrons induced by closing the gap overcomes the cost of the relaxation energy of reconfiguration, a nearly gapless TSS will occur. Exemplified the MnBi$_{2}$Te$_{4}$ family, we demonstrate the above mechanism by applying the modern theory of doping implemented into density functional theory (DFT) calculations. The calculated ground states with gapless and gapped TSS spectra for MnBi$_{2}$Te$_{4}$ family yield nice agreements with the experimental observations. Our work also shed light on the design principles of MTI with magnetic gaps, revealing the critical role of doping effects in understanding the magnetism-induced topological phase transitions.

\emph{Model Study}.\rule[3pt]{0.4cm}{0.02em}Koopmans' theorem uses the Hartree-Fock method for the approximation of single-particle orbital energy, stating that the first ionization energy is equal to the highest occupied orbital energy \cite{Ref.29,Ref.30,Ref.31}. It can be easily generalized to calculate the energy changes when electrons are added to or removed from a \emph{N}-electron system, e.g., $E(N+1)-E(N)=\varepsilon_{N+1}$ for \emph{n}-type doping, where \emph{E}(\emph{N}) and $\varepsilon_{N+1}$ denote the total energy of the undoped system and the single-particle energy eigenvalue for electron addition, respectively. For a metallic system, one can define the total energy difference upon doping \emph{E}$_{D}$ ($\delta$) as a function of the continuous change of the occupation number $\delta$, and apply the generalized Koopmans' theorm as follows:
\begin{equation}
\begin{aligned}
  E_D(\delta)&= E(N+\delta) - E(N)=\varepsilon_F \cdot \delta
\end{aligned}
\end{equation}
where $\varepsilon_F$ denotes the Fermi level. Besides ionization and affinity energy calculations, the obtained linear relationship between the total energy and occupation number is widely used to benchmark the self-interaction error of DFT \cite{Ref.30,Ref.31,Ref.32}.

\begin{figure}[thb!]
\centering
\includegraphics[width=0.40\textwidth]{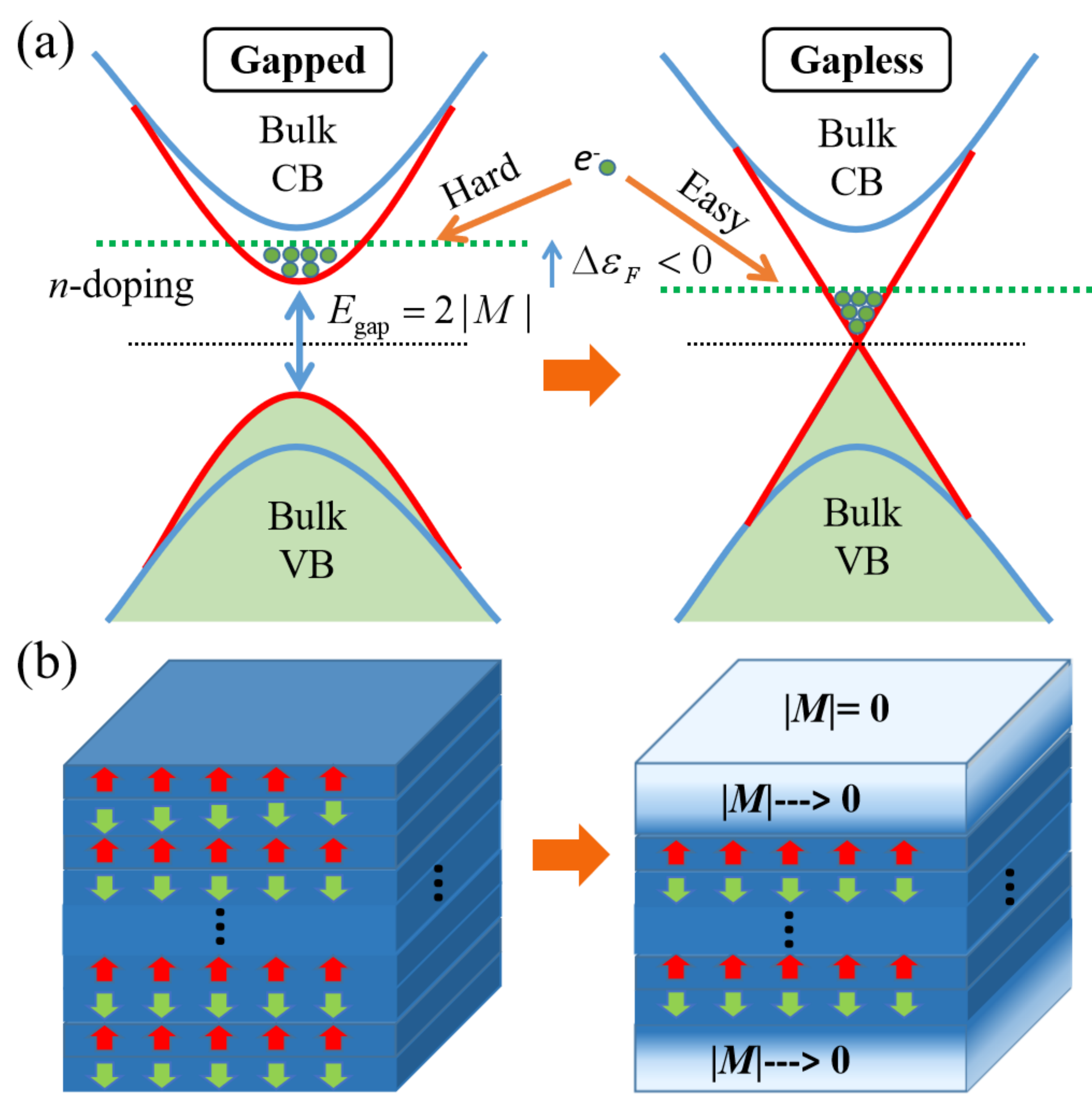}%
\caption{\label{fig1} \raggedright (a) The sketch of the self-doped MTI with gapped and gapless topological surface state (TSS), showing that in doped MTI a phase transition from gapped to gapless TSS is likely to occur. The surface (bulk) bands are represented by the red (blue) line. The black and green dash line show the charge neutrality and the Fermi level for \emph{n}-type doping, respectively. (b) Corresponding reconstruction of the magnetic or geometric configuration of a doped A-AFM topological insulator. The surface tends to form a smooth domain with the effective out-of-plane magnetization $|\emph{M}|$$\rightarrow$0.}
\end{figure}

One unique feature of the MTI is the multiple phases with nearly degenerate energies but distinct surface states. The TSS dominates the states near $\varepsilon_F$ and thus effectively couples the doping effect; so the doped MTI may adopt a specific topological surface state that is favorable in energy. As schematically shown in Fig. \ref{fig1}(a), we consider two distinct topological surface states, one has a magnetic gap and the other is gapless for whatever reason. Applying Eq. (1) to the \emph{n}-doped gapless and gapped systems, respectively, the difference of the doping energy \emph{E}$_{D}$ ($\delta$) between them depends on the difference of their $\varepsilon_F$, which is expressed as follows:
\begin{equation}
\begin{aligned}
  E_{D}^{gapless}-E_{D}^{gapped}&= \bigtriangleup \varepsilon_F \cdot \delta \approx -|M| \cdot \delta
\end{aligned}
\end{equation}
When $\delta$ electrons are occupying the surface states, Eq. (2) indicates that $\bigtriangleup$$\varepsilon_F$ (i.e., $\varepsilon_F^{gapless} - \varepsilon_F^{gapped}$) is less than zero, and approaches to the value of $-|\emph{M}|$ as $\delta$ down to zero. Therefore, injecting a few electrons into the gapless system is easier than the gapped system by earning the doping energy $|\emph{M}|\cdot\delta$. For the case \emph{p}-type doping, a gapless surface state is still favored for doped MTI (see Fig. S1 \cite{Ref.33}). Overall, self-doping tends to promote the transition from gapped TSS to gapless TSS in a MTI.

For a MTI whose charge-neutral ground state is gapped by magnetism, whether such transition occurs or not depends on the competition between the energy gain from Eq. (2) and the relaxation energy cost to change the crystal geometry or magnetic configuration. If the relaxation energy is small enough, e.g., the magnetic anisotropy energy (MAE, defined as $\emph{E}_{x}-\emph{E}_{z}$, where \emph{E}$_{x}$ and \emph{E}$_{z}$ are the total energies with in-plane and out-of-plane magnetization, respectively), the system naturally tends to reduce its gap to earn the doping energy \emph{E}$_{D}$ ($\delta$). Taking MnBi$_{2}$Te$_{4}$ as an example, whose bulk ground state is reported to be A-type AFM with the intralayer FM spin alignment along the \emph{z} direction. According to our DFT calculation as well as the results from the literature, $|\emph{M}|$ is around 40$\sim$50 meV \cite{Ref.6}. In comparison, for the paramagnetic phase or AFM phase with in-plane spin orientation, $|\emph{M}|$ is almost vanishing, resulting in a nearly gapless TSS. According to Eq. (2), only a fraction of electron per unit cell (u.c.) at the surface is capable to earn enough \emph{E}$_{D}$ ($\delta$) that overwhelms the MAE (0.4 meV/Mn \cite{Ref.10} times the thickness of the domain). Since the synthesized MnBi$_{2}$Te$_{4}$ is typically \emph{n}-type self-doped, the spin reorientation from out-of-plane to in-plane magnetic moment has the legitimate driven force to take place, giving rise to a nearly gapless TSS \cite{Ref.10}. In the following, we verify the above model in the self-doped MTI MnBi$_{2}$Te$_{4}$/(Bi$_{2}$Te$_{3}$)$_{n}$ family by using comprehensive DFT calculations.

\emph{DFT Results}.\rule[3pt]{0.4cm}{0.02em}For a specific doped-MTI, the transition from gapped to gapless TSS could be attributed to various material-dependent reasons, e.g., surface magnetic or geometric reconstruction \cite{Ref.10,Ref.12,Ref.18,Ref.19,Ref.20,Ref.21}. In principle, all these reasons are dictated by the fundamental energy-lowering principle from Koopmans' theorm discussed above. Without loss of generality, we next choose magnetic anisotropy (MA) transition at the surface of self-doped MnBi$_{2}$Te$_{4}$/(Bi$_{2}$Te$_{3}$)$_{n}$ family as one of most possible factor of nearly gapless TSS. As shown in Fig. \ref{fig1}(b), magnetic moments align out-of-plane in the bulk and turn to in-plane at the top/bottom surface, connected by a magnetic domain wall. If the domain wall is thick enough, the energy cost of the domain wall is negligible. Thus, we first choose a 6-SL MnBi$_{2}$Te$_{4}$ slab to simulate the MAE effects and the transition between different TSS. The calculated band gap of the 6-SL slab with in-plane magnetization is only about 1.6 meV, indicating that the hybridization between the top and bottom surface is negligible. To validate our results, the doping effect implemented into the DFT calculations is realized by two means. One is the virtual crystal approximation (VCA) approach, which considers a symmetry-preserved primitive cell composed of ¡°virtual¡± atoms \cite{Ref.34}. The other one is the supercell approach with two types of antisite defects to simulate the chemical doping in MnBi$_{2}$Te$_{4}$. More details of DFT methods are provided in Supplemental Materials \cite{Ref.33}.

\begin{figure}[thb!]
\centering
\includegraphics[width=0.45\textwidth]{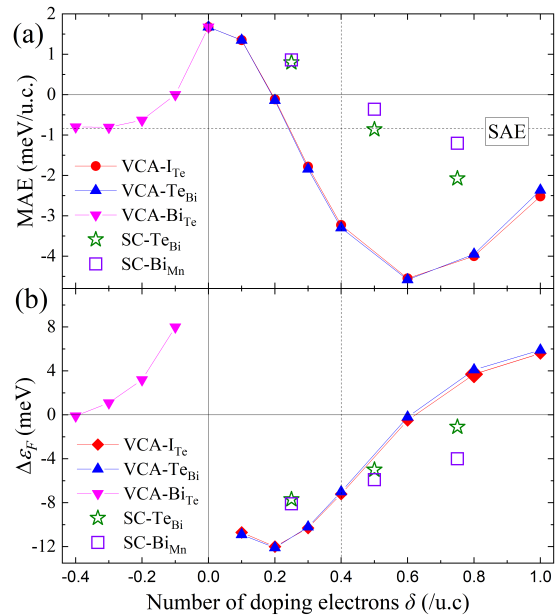}%
\caption{\label{fig2} \raggedright (a) MAE and (b) $\bigtriangleup$$\varepsilon_F$ of 6-SL MnBi$_{2}$Te$_{4}$ slabs as a function of the doping electrons $\delta$, calculated by VCA and supercell approaches. For VCA approach, the substitution donor defect I$_{Te}$, Te$_{Bi}$ and the acceptor defect Bi$_{Te}$ are considered. For supercell approach, Te$_{Bi}$ and Bi$_{Mn}$ substitutions are considered. The dotted horizontal line in (a) represents the SAE of the slab. The dotted vertical line corresponds to $\delta$ = 0.4, equivalent to the doping concentration of 3.0$\times$10$^{20}$cm$^{-3}$.}
\end{figure}

We first study the ground state of doped MnBi$_{2}$Te$_{4}$ by VCA method. The calculated MAE as a function of the doping electrons $\delta$ in 6-SLs slab is shown in Fig. \ref{fig2}(a). The unit of the MAE and $\delta$ are defiend as meV/u.c. and /u.c., respectively (note that the u.c. stands for a 2D slab with finite thickness). The MAE mainly contains two factors, i.e., magnetocrystalline anisotropy originated from spin-orbit coupling, and shape anisotropy energy (SAE) from magnetic dipole-dipole interaction, which always prefers in-plane magnetization for thin slabs. The SAE, denoted by the dotted horizontal line, is taken as -0.14 meV/Mn \cite{Ref.35}. For the undoped system, the calculated magnetocrystalline anisotropy energy is 0.42 meV/Mn, which is close to previous results \cite{Ref.10}. The substitution donor defect I$_{Te}$, Te$_{Bi}$ and the acceptor defect Bi$_{Te}$ are considered. As shown in Fig. \ref{fig2}(a), no matter \emph{p}-type or \emph{n}-type doping, a ground state with out-of-plane magnetic moment is only preferred within a narrow doping range. For n-type doping, the MAE drops below zero when $\delta$ is larger than 0.2, while upturns after 0.6 electrons doping. This is because more bulk states instead of surface states are occupied, reducing the effective doping concentration. In addition, for all three types of doping, the MAE for the doped MnBi$_{2}$Te$_{4}$ slabs are very close. This indicates that the response of MAE upon doping mostly comes from the doping electrons, rather than the specific potential of an individual defect.

We also calculate $\bigtriangleup$$\varepsilon_F$ between the gapless (in-plane magnetization) and gapped system (out-of-plane magnetization), as shown in Fig. \ref{fig2}(b). Here, the Fermi energys are calculated related to the atom core level energy for the comparison between different systems. We find that $\bigtriangleup$$\varepsilon_F$ is positive for \emph{p}-type doping, and negative for light \emph{n}-type doping ($\delta$ $<$ 0.6). When $\delta$ is larger than 0.6, $\bigtriangleup$$\varepsilon_F$ becomes to be positive, exactly corresponding to the upturn of MAE shown in Fig. \ref{fig2}(a). Such consistency perfectly fulfills our model Eq. (2) derived from generalized Koopmans' theorm, implying that $\bigtriangleup$$\varepsilon_F$ reflects the derivative of MAE to the doping concentration.

To further validate our model, we next consider the chemical doping effect by applying the supercell approach. It takes into account the local disorder effect, which could modulate the electron structure \cite{Ref.36}. We construct a 168-atom supercell of MnBi$_{2}$Te$_{4}$ to simulate the chemical doping. One, two and three Te$_{Bi}$ and Bi$_{Mn}$ substitutions ($\delta$ = 0.25, 0.50 and 0.75, respectively), which are two of the most possible donor defects in MnBi$_{2}$Te$_{4}$ \cite{Ref.27,Ref.37}, are considered with nearly homogeneous distribution. As shown in Fig. \ref{fig2}(a), the MAE of MnBi$_{2}$Te$_{4}$ with both Te$_{Bi}$ and Bi$_{Mn}$ defects decrease with increasing $\delta$, and become negative with $\delta$ = 0.5 indicating the transition to in-plane magnetization. In accordance, $\bigtriangleup$$\varepsilon_F$ for Te$_{Bi}$ and Bi$_{Mn}$ defects are all negative as shown in Fig. \ref{fig2}(b). Such qualitatively consistent results compared to the VCA method indicates the validity of our model for the MTI materials such as MnBi$_{2}$Te$_{4}$.

\begin{figure}[thb!]
\centering
\includegraphics[width=0.45\textwidth]{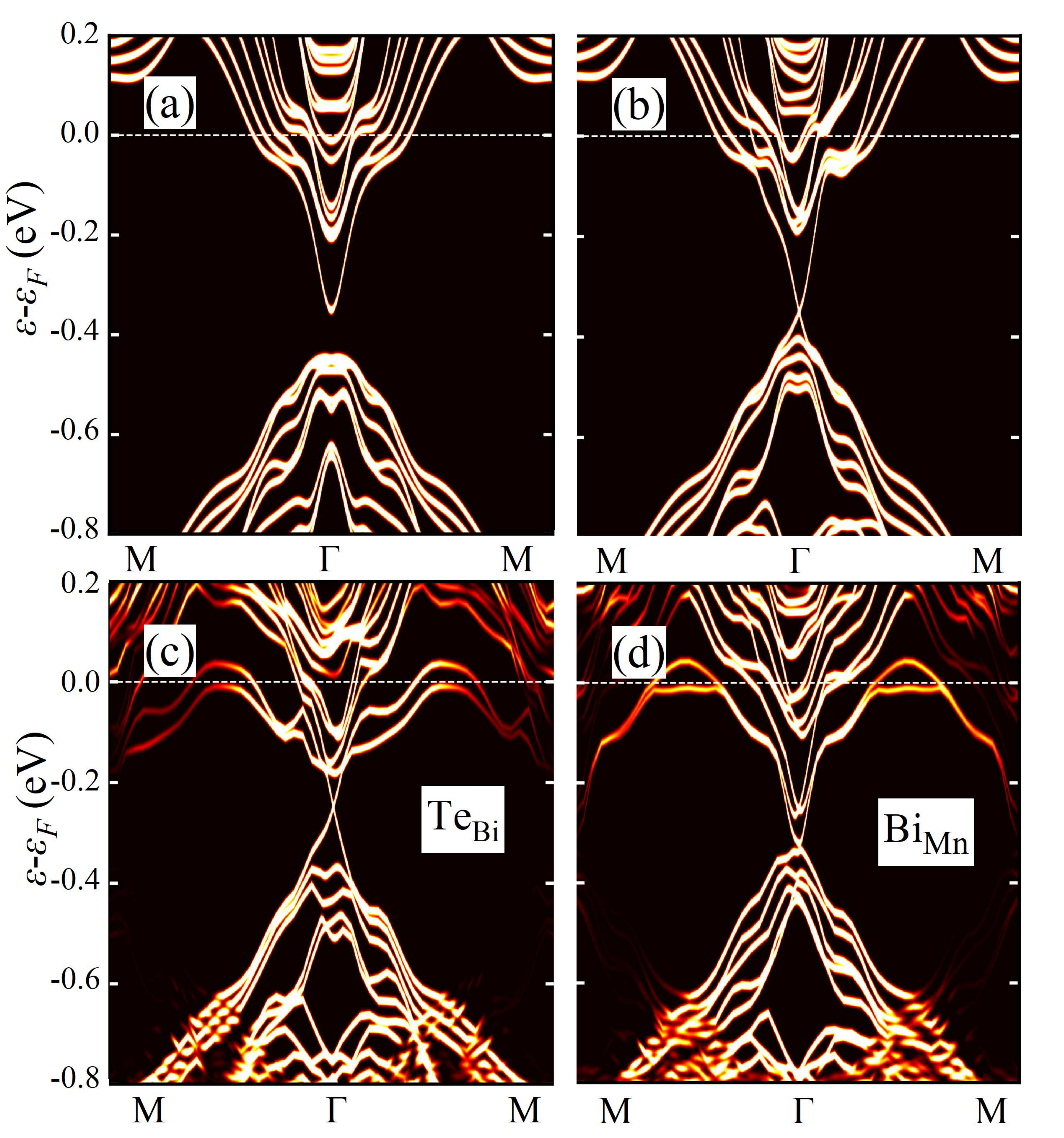}%
\caption{\label{fig3} \raggedright (a-b) Band structures of 6-SL MnBi$_{2}$Te$_{4}$ slabs with 0.4/u.c. electrons doping induced by Te$_{Bi}$ defects calculated by VCA approach for (a) out-of-plane and (b) in-plane magnetization. (c-d) The effective band structure of the ground states with in-plane magnetization of 6-SL MnBi$_{2}$Te$_{4}$ slabs with 0.5/u.c. electrons doping induced by (c) Te$_{Bi}$ and (d) Bi$_{Mn}$ defects calculated by supercell approach. The dash lines mark the Fermi level of the doped sample.}
\end{figure}

In experiment, all the nearly gapless TSS of MnBi$_{2}$Te$_{4}$, observed via ARPES \cite{Ref.10,Ref.11,Ref.12,Ref.13}, possess the Dirac cone located at about 280 meV below the experimental Fermi level, indicating their doping concentrations. In order to benchmark the measured doping level, we calculate the band structures of doped 6-SL MnBi$_{2}$Te$_{4}$ with the in-plane magnetization as the ground state. Both of the VCA (Te$_{Bi}$ defect) and the supercell (Te$_{Bi}$ and Bi$_{Mn}$ defects) approaches are applied, with the doping electrons per u.c. $\delta$ = 0.4 (corresponding to the doping concentration of 3.0$\times$10$^{20}$cm$^{-3}$, close to the experimental doping concentration in MnBi$_{2}$Te$_{4}$ \cite{Ref.26}) and 0.5, respectively. As shown in Fig. \ref{fig3}(a) and \ref{fig3}(b), out-of-plane and in-plane magnetization give rise to magnetic gap and gapless TSS, respectively, while the charge neutral point of the Dirac cone is 325 meV below $\varepsilon_F$, close to the experimental value. Therefore, MA transition induced by self-doping could occur at the surface of experimental MnBi$_{2}$Te$_{4}$ samples. For the supercell approach, similar conclusion stands. We extract the ¡°effective band structure¡± (EBS) \cite{Ref.38,Ref.39} to unfold the sophisticated \emph{E}-\emph{k} spaghetti within a supercell Brillouin zone (BZ) into the spectrum density within the primitive BZ. Fig. \ref{fig3}(c) and \ref{fig3}(d) show the EBS for the ground states with in-plane magnetization of MnBi$_{2}$Te$_{4}$ with Te$_{Bi}$ and Bi$_{Mn}$ defects ($\delta$ = 0.5), respectively. Both EBS spectra show nearly gapless feature. In details, there is an unambiguous band crossing point located at -230 meV for non-magnetic Te$_{Bi}$ defects, while the charge neutral point with deformation around -300 meV for magnetic Bi$_{Mn}$ defects. Overall, our DFT calculations show that self-doping indeed results in the nearly gapless spectra in doped MnBi$_{2}$Te$_{4}$ samples, consistent with our theoretical model from the generalized Koopmans' theorem.

\begin{figure}[thb!]
\centering
\includegraphics[width=0.45\textwidth]{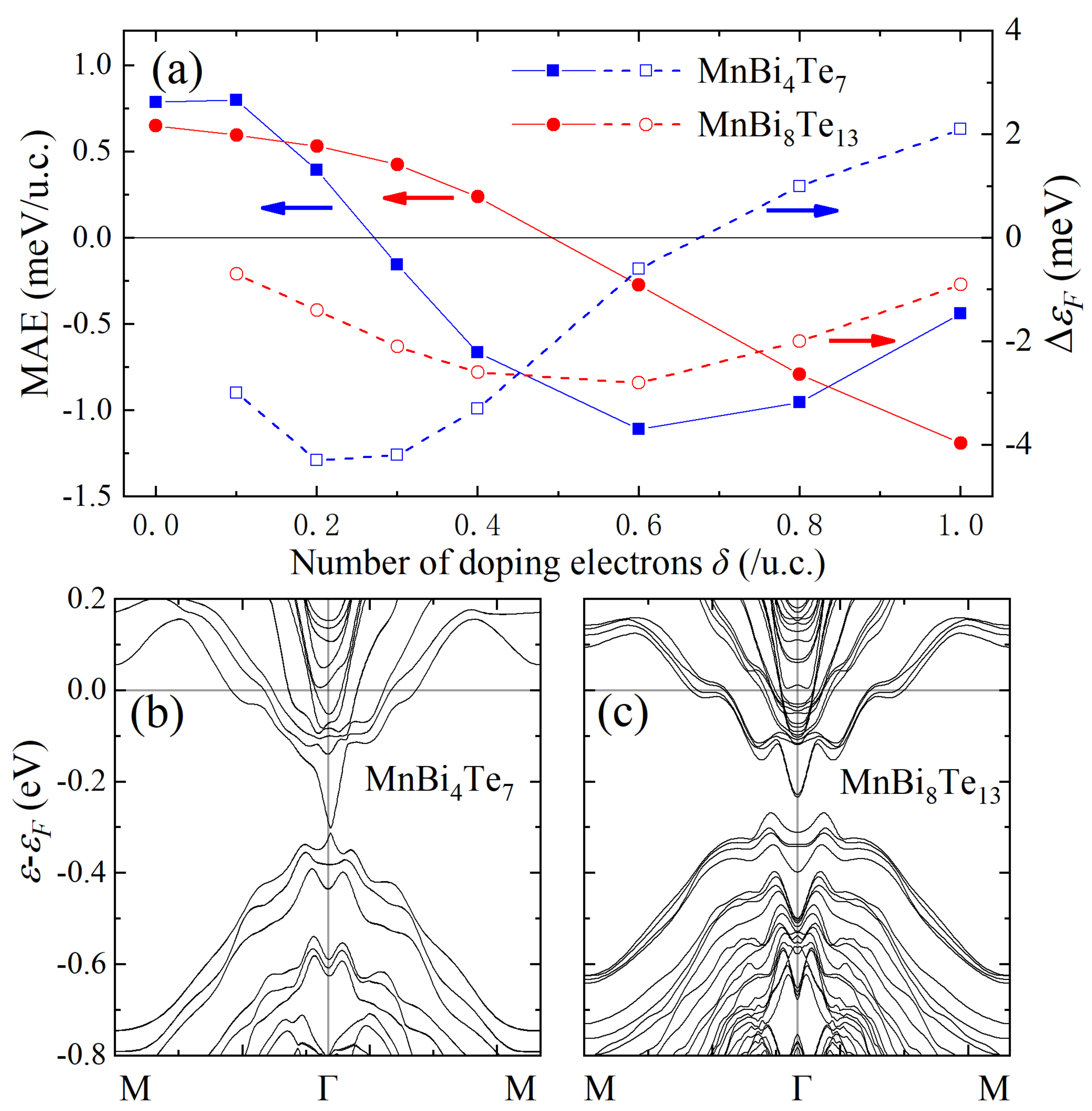}%
\caption{\label{fig4} \raggedright (a) MAE and $\bigtriangleup$$\varepsilon_F$ of doped MnBi$_{4}$Te$_{7}$ and MnBi$_{8}$Te$_{13}$ slabs with Te$_{Bi}$ defects calculated by VCA approach. The solid line and dashed line represent MAE and $\bigtriangleup$$\varepsilon_F$, respectively. (b-c) The band structure of the ground states for (b) MnBi$_{4}$Te$_{7}$ slab with in-plane magnetization and (c) MnBi$_{8}$Te$_{13}$ slab with out-of-plane magnetization with 0.4/u.c. electrons doping.}
\end{figure}

Besides MnBi$_{2}$Te$_{4}$, MnBi$_{2}$Te$_{4}$/(Bi$_{2}$Te$_{3}$)$_{n}$ vdW MTI family with \emph{n} = 1, 2 and 3 Bi$_{2}$Te$_{3}$ quintuple-layers (QL) were also grown as single crystals and characterized by ARPES \cite{Ref.23,Ref.40,Ref.41,Ref.42,Ref.43,Ref.44,Ref.45,Ref.46}. While the nearly gapless TSS was observed at the MnBi$_{2}$Te$_{4}$-termination of MnBi$_{4}$Te$_{7}$ \cite{Ref.10,Ref.15,Ref.22,Ref.23,Ref.24}, surprisingly, a magnetic gap about 28 meV was verified at the MnBi$_{2}$Te$_{4}$-termination of MnBi$_{8}$Te$_{13}$ via APRES measurements \cite{Ref.25}. To validate our theory, we perform DFT calculations for doped MnBi$_{4}$Te$_{7}$ slab stacked as SL-QL-SL-QL-SL-QL-SL (7 vdW layers) and doped MnBi$_{8}$Te$_{13}$ slab stacked as SL-3(QL)-SL-3(QL)-SL (9 vdW layers) via VCA approach with Te$_{Bi}$ defects. Fig. \ref{fig4}(a) shows the MAE of the doped MnBi$_{4}$Te$_{7}$ (MnBi$_{8}$Te$_{13}$) denoted by the solid blue square (solid red circle), and $\bigtriangleup$$\varepsilon_F$ between the doped MnBi$_{4}$Te$_{7}$ (MnBi$_{8}$Te$_{13}$) with in-plane and out-of-plane magnetization denoted by the hollow blue square (hollow red circle). In light doping range ($\delta$ $<$ 0.6), the MAE decreases for both MnBi$_{4}$Te$_{7}$ and MnBi$_{8}$Te$_{13}$ slabs with increasing $\delta$, accompanied with negative $\bigtriangleup$$\varepsilon_F$, consistent with Koopmans' theorem. However, comparing with MnBi$_{4}$Te$_{7}$, the decline of MAE upon electron doping in MnBi$_{8}$Te$_{13}$ is slower. For $\delta$ = 0.4, the MA transition only occurs for doped MnBi$_{4}$Te$_{7}$, while the doped MnBi$_{8}$Te$_{13}$ keeps the out-of-plane magnetization as the ground state. Consequently, as shown in Fig. 4(c), the ground state of the doped MnBi$_{4}$Te$_{7}$ exhibits nearly gapless TSS located at -310 meV, which is similar to that of MnBi$_{2}$Te$_{4}$. In contrast, the TSS of doped MnBi$_{8}$Te$_{13}$ has a magnetic gap as shown in Fig. \ref{fig4}(d), where the conduction band minimum is around -235 meV, also close to the experimental value (about -200 meV) \cite{Ref.25}.

To explain the distinct behavior of TSS of MnBi$_{4}$Te$_{7}$ and MnBi$_{8}$Te$_{13}$, it is straightforward to attribute to their AFM and FM magnetic configurations. However, we note that despite the AFM nature of MnBi$_{4}$Te$_{7}$, the penetration of the TSS, which is mainly captured by ARPES, is less than two vdW layers, similar to that of MnBi$_{8}$Te$_{13}$. Here we provide an alternative explanation. As shown in Fig. \ref{fig3}(c) and \ref{fig3}(d), the conduction band of the bulk states are partially occupied by the doping electrons. As mentioned in the model, only the doping electrons occupying the TSS contribute to the energy gain and the transition. On the other hand, the Dirac cone of Bi$_{2}$Te$_{3}$ is deeper than the charge neutral point of MnBi$_{2}$Te$_{4}$ \cite{Ref.20}. Therefore, more Bi$_{2}$Te$_{3}$ vdW layers in MnBi$_{8}$Te$_{13}$ than that in MnBi$_{4}$Te$_{7}$ result in more low-energy bulk conduction bands occupied by self-doping electrons, thus reducing the effective $\delta$ for MA transition to a gapless TSS.

\emph{Discussion}.\rule[3pt]{0.4cm}{0.02em}Our theory based on the generalized Koopmans' theorem reaches the conclusion that a doped MTI tends to have a gapless TSS to gain the single-particle energy of the surface state, thus fundamentally explaining the nearly gapless TSS of MnBi$_{2}$Te$_{4}$ family observed by multiple ARPES measurements \cite{Ref.10,Ref.11,Ref.12,Ref.13,Ref.14,Ref.15,Ref.20,Ref.24,Ref.26,Ref.41}. On the other hand, it can also explain the observation of QAHE observed in a 5-SL MnBi$_{2}$Te$_{4}$ slab \cite{Ref.47}, which implies that the finite magnetic gap overwhelms the hybridization gap. Note that to observe quantized anormalous Hall resistance, a large gate voltage (\emph{V}$_{g}$ = 200 V) is required to tune the $\varepsilon_F$ to the charge neutral point. From the perspective of our model, \emph{V}$_{g}$ removes the self-doping effect, maintaining the ground state with a magnetic gap \cite{Ref.48}. Thus, we suggest that the topological gap (0.64 meV, \cite{Ref.47}) hinted by the realization temperature is smaller than the magnetic surface gap of the charge-neutral MnBi$_{2}$Te$_{4}$ because of the inhomogeneous band alignment of the grown sample.

In this sense, our work also shed light on several design principles of MTI with magnetic gaps. According to Eq. (2), one can either reduce the doping energy by looking for materials with large formation energy of charged defects, which usually requires strong chemical bonding and stable structural networks, or increase the relaxation energy, e.g., looking for materials with large MAE and exchange energy. Furthermore, our work not only demystifies the deviation between the theoretical predictions and experimental measurements on the existence of the magnetic gap in MnBi$_{2}$Te$_{4}$ family, but also uncovers the essential role of the previously overlooked doping effects in understanding certain delicate interplay between magnetism and topology.

This work was supported by National Key R$\&$D Program of China under Grant Nos. 2020YFA0308900 and 2019YFA0704900, Guangdong Innovative and Entrepreneurial Research Team Program under Grant No. 2017ZT07C062, Guangdong Provincial Key Laboratory for Computational Science and Material Design under Grant No. 2019B030301001, the Shenzhen Science and Technology Program (Grant No.KQTD20190929173815000) and Center for Computational Science and Engineering of Southern University of Science and Technology.


\begin{thebibliography}{19}%
\makeatletter
\providecommand \@ifxundefined [1]{%
 \@ifx{#1\undefined}
}%
\providecommand \@ifnum [1]{%
 \ifnum #1\expandafter \@firstoftwo
 \else \expandafter \@secondoftwo
 \fi
}%
\providecommand \@ifx [1]{%
 \ifx #1\expandafter \@firstoftwo
 \else \expandafter \@secondoftwo
 \fi
}%
\providecommand \natexlab [1]{#1}%
\providecommand \enquote  [1]{``#1''}%
\providecommand \bibnamefont  [1]{#1}%
\providecommand \bibfnamefont [1]{#1}%
\providecommand \citenamefont [1]{#1}%
\providecommand \href@noop [0]{\@secondoftwo}%
\providecommand \href [0]{\begingroup \@sanitize@url \@href}%
\providecommand \@href[1]{\@@startlink{#1}\@@href}%
\providecommand \@@href[1]{\endgroup#1\@@endlink}%
\providecommand \@sanitize@url [0]{\catcode `\\12\catcode `\$12\catcode
  `\&12\catcode `\#12\catcode `\^12\catcode `\_12\catcode `\%12\relax}%
\providecommand \@@startlink[1]{}%
\providecommand \@@endlink[0]{}%
\providecommand \url  [0]{\begingroup\@sanitize@url \@url }%
\providecommand \@url [1]{\endgroup\@href {#1}{\urlprefix }}%
\providecommand \urlprefix  [0]{URL }%
\providecommand \Eprint [0]{\href }%
\providecommand \doibase [0]{http://dx.doi.org/}%
\providecommand \selectlanguage [0]{\@gobble}%
\providecommand \bibinfo  [0]{\@secondoftwo}%
\providecommand \bibfield  [0]{\@secondoftwo}%
\providecommand \translation [1]{[#1]}%
\providecommand \BibitemOpen [0]{}%
\providecommand \bibitemStop [0]{}%
\providecommand \bibitemNoStop [0]{.\EOS\space}%
\providecommand \EOS [0]{\spacefactor3000\relax}%
\providecommand \BibitemShut  [1]{\csname bibitem#1\endcsname}%

\bibitem{Ref.1} L. $\check{S}$mejkal, Y. Mokrousov, B. Yan, and A. H. MacDonald, Nat. Phys. \textbf{14}, 242 (2018).
\bibitem{Ref.2} Y. Tokura, K. Yasuda, and A. Tsukazaki, Nat. Rev. Phys. \textbf{1}, 126 (2019).
\bibitem{Ref.3} C. X. Liu, X. L. Qi, X. Dai, Z. Fang, and S. C. Zhang, Phys. Rev. Lett. \textbf{101}, 146802 (2008).
\bibitem{Ref.4} R. Yu, W. Zhang, H.-J. Zhang, S.-C. Zhang, X. Dai, and Z. Fang, Science \textbf{329} 61 (2010).
\bibitem{Ref.5} C.-Z. Chang et al., Science \textbf{340}, 167 (2013).
\bibitem{Ref.6} M. M. Otrokov et al., Nature \textbf{576}, 416 (2019).
\bibitem{Ref.7} D. Zhang, M. Shi, T. Zhu, D. Xing, H. Zhang, and J. Wang, Phys. Rev. Lett. \textbf{122}, 206401 (2019).
\bibitem{Ref.8} J. Li, Y. Li, S. Du, Z. Wang, B.-L. Gu, S.-C. Zhang, K. He, W. Duan, and Y. Xu, Sci. Adv. \textbf{5}, eaaw5685 (2019).
\bibitem{Ref.9} R. C. Vidal et al., Phys. Rev. B \textbf{100}, 121104(R) (2019).
\bibitem{Ref.10} Y.-J. Hao et al., Phys. Rev. X \textbf{9}, 041038 (2019).
\bibitem{Ref.11} Y. J. Chen et al., Phys. Rev. X \textbf{9}, 041040 (2019).
\bibitem{Ref.12} P. Swatek, Y. Wu, L.-L. Wang, K. Lee, B. Schrunk, J. Yan, and A. Kaminski, Phys. Rev. B \textbf{101}, 161109(R) (2020).
\bibitem{Ref.13} D. Nevola, H. X. Li, J. Q. Yan, R. G. Moore, H. N. Lee, H. Miao, and P. D. Johnson, Phys. Rev. Lett. \textbf{125}, 117205 (2020).
\bibitem{Ref.14} Z. Liang et al., Phys. Rev. B \textbf{102}, 161115(R) (2020).
\bibitem{Ref.15} H. Li et al., Phys. Rev. X \textbf{9}, 041039 (2019).
\bibitem{Ref.16} B. Chen et al., Nat. Commun. \textbf{10}, 4469 (2019).
\bibitem{Ref.17} Y. Gong et al., Chin. Phys. Lett. \textbf{36}, 076801 (2019).
\bibitem{Ref.18} H.-P. Sun et al., Phys. Rev. B \textbf{102}, 241406(R) (2020).
\bibitem{Ref.19} F. Hou et al., ACS Nano \textbf{14}, 11262 (2020).
\bibitem{Ref.20} Y. Yuan et al., Nano Lett. \textbf{20}, 3271 (2020).
\bibitem{Ref.21} A. M. Shikin et al., Sci. Rep. \textbf{10}, 13226 (2020).
\bibitem{Ref.22} X. Wu et al., Phys. Rev. X \textbf{10}, 031013 (2020).
\bibitem{Ref.23} L. Ding, C. Hu, F. Ye, E. Feng, N. Ni, and H. Cao, Phys. Rev. B \textbf{101}, 020412(R) (2020).
\bibitem{Ref.24} Y. Hu et al., Phys. Rev. B \textbf{101}, 161113(R) (2020).
\bibitem{Ref.25} R. Lu et al., arXiv:2009.04140 (2020).
\bibitem{Ref.26} J. Q. Yan et al., Phys. Rev. Materials \textbf{3}, 064202 (2019).
\bibitem{Ref.27} Z. Huang, M.-H. Du, J. Yan, and W. Wu, Phys. Rev. Materials \textbf{4}, 121202(R) (2020).
\bibitem{Ref.28} A. Zeugner et al., Chem. Mater. \textbf{31}, 2795 (2019).
\bibitem{Ref.29} T. C. Koopmans, Physica (Utrecht)  \textbf{1}, 104 (1934).
\bibitem{Ref.30} J. P. Perdew, R. G. Parr, M. Levy, and J. L. Balduz, Phys. Rev. Lett. \textbf{49}, 1691 (1982).
\bibitem{Ref.31} S. Lany and A. Zunger, Phys. Rev. B \textbf{80}, 085202 (2009).
\bibitem{Ref.32} Q. Liu, Q. Yao, Z. A. Kelly, C. M. Pasco, T. M. McQueen, S. Lany, and A. Zunger, Phys. Rev. Lett. \textbf{121}, 186402 (2018).
\bibitem{Ref.33} See Supplemental Materials for the model for $\emph{p}$-type doped magnetic topological insulator, DFT methods and the supplementary figures.
\bibitem{Ref.34} L. Bellaiche and D. Vanderbilt, Phys. Rev. B \textbf{61}, 7877 (2000).
\bibitem{Ref.35} F. Xue, Z. Wang, Y. Hou, L. Gu, and R. Wu, Phys. Rev. B \textbf{101}, 184426 (2020).
\bibitem{Ref.36} J. Shen et al., Phys. Rev. Lett. \textbf{125}, 086602 (2020).
\bibitem{Ref.37} M. H. Du, J. Yan, V. R. Cooper, and M. Eisenbach, Adv. Funct. Mater., 2006516 (2020).
\bibitem{Ref.38} V. Popescu and A. Zunger, Phys. Rev. Lett. \textbf{104}, 236403 (2010).
\bibitem{Ref.39} P. V. C. Medeiros, S. Stafstr$\ddot{o}$m, and J. Bj$\ddot{o}$rk, Phys. Rev. B \textbf{89}, 041407(R) (2014).
\bibitem{Ref.40} J. Wu et al., Sci. Adv. \textbf{5}, eaax9989 (2019).
\bibitem{Ref.41} C. Hu et al., Nat. Commun. \textbf{11}, 97 (2020).
\bibitem{Ref.42} J. Q. Yan, Y. H. Liu, D. S. Parker, Y. Wu, A. A. Aczel, M. Matsuda, M. A. McGuire, and B. C. Sales, Phys. Rev. Materials \textbf{4}, 054202 (2020).
\bibitem{Ref.43} R. C. Vidal et al., Phys. Rev. X \textbf{9}, 041065 (2019).
\bibitem{Ref.44} Z. S. Aliev et al., Journal of Alloys and Compounds \textbf{789}, 443 (2019).
\bibitem{Ref.45} M. Z. Shi, B. Lei, C. S. Zhu, D. H. Ma, J. H. Cui, Z. L. Sun, J. J. Ying, and X. H. Chen, Phys. Rev. B \textbf{100}, 155144 (2019).
\bibitem{Ref.46} C. Hu et al., Sci. Adv. \textbf{6}, eaba4275 (2020).
\bibitem{Ref.47} Y. Deng, Y.jun Yu, M. Z. Shi, Z. Guo, Z. Xu, J. Wang, X. H. Chen, and Y. Zhang, Science \textbf{367}, 895 (2020).
\bibitem{Ref.48} We note that certain pinning of the in-plane domain wall may still occur and thus reduce the magnetic gap and the critical temperature of QAHE.



\end{thebibliography}
\end{document}